# Regenerative Soot-VII: Population inversion of He I and Ne I in regenerative sooting discharges

**Sajid Hussain[1], A Aleem[1], Rahila Khalid[1], S D Khan[1], A Ellahi[2], S A Janjua[1] and Shoaib Ahmad[1,2,,*3]**

[1] PINSTECH, P O Nilore, Islamabad, Pakistan
[2] CASP, Government College University, Lahore, Pakistan
[3] National Centre for Physics, Quaid-i-Azam University Campus, Shahdara Valley, Islamabad, 44000, Pakistan

*Email: sahmad.ncp@gmail.com

**Abstract**

The mechanisms for population inversion among the excited states of He I and Ne I in regenerative sooting discharges have been investigated as a function of the geometry and the physical parameters of the graphite hollow cathode sources. The effect of the state of sooting of the hollow cathode, the gas pressure and the role of the cycling of $i_{dis}$ on the population inversion of the excited He I and Ne I levels have been investigated.

## 1. Introduction

Sooting discharges created in graphite hollow cathodes have been investigated in our laboratory and results published on different aspects of such discharges. A review article [1] has discussed and described the state of understanding of the physical processes involved in the emission spectroscopy and mass spectrometry of regenerative sooting discharges up to 2002; the present study is expected to add new information especially with regard to the excited states of the discharge gases, i.e. He and Ne. The sooting discharge is created and sustained in graphite hollow cathodes whose surface is first sputtered by the energetic discharge species and later on gets covered by layers of carbon clusters. The initial and most significant stage in the production of regenerative soot depends on the initiation of the glow discharge of the noble gas by two well-defined electron energy regimes [2]: high energy, 10 eV, electrons near the cathode surface and low energy ones, 1 eV, in the positive column. The lower energy electrons are expected to have Maxwellian velocity distributions; therefore, one can estimate the electron temperature of the positive column of the glow discharge and approximate it with the excitation temperature of the support gas by using the most pronounced emission lines with the same lower levels. However, the presence of the higher energy electrons ( 10 eV) near the electron-emitting cathode and the role of the sooted surface of the hollow cathode as well as the inter-and intra-species collisions in the positive column of the discharge provide additional sources of excitation besides the low energy electrons ( 1 eV).

Population inversion of the excited level densities in sooting discharges can serve as a parameter for the evaluation of the discharge characteristics. Emission spectroscopy of the C containing He I and Ne I discharges has revealed that these discharges are composed of atomic and ionic species that are generally not in LTE. The discharge current cycling between the minimum and maximum values provides important clues to the level densities of the respective transitions. In a recent study [3], we focused on the excited and ionized states of $C_1$ under similar experimental conditions, where we pointed out hysteresis in the level densities as a function of the cycling of $i_{dis}$ between min → max → min. Since $C_1$ is created as a by-product of the sooting discharge, one can argue that it may not constitute conclusive evidence of the mechanism of hysteresis. Therefore, the present series of experiments has targeted the noble gas's atomic species emission lines to probe the hysteresis of higher excited levels as well as to provide conclusive evidence of population inversion. One of the criteria for population inversion for a three-level atom [4] is that the middle level must have a faster decay time to the ground (or a lower) level as compared with the upper-middle level, i.e. $\tau_{upper\text{-}middle} > \tau_{middle\text{-}lower}$. This criterion requires the complete decay scheme for the levels considered. However, the calculated level densities from the measured intensities of various transitions can provide unambiguous proof of population inversion and the parameter involved is $(N_u \cdot g_l / N_l \cdot g_u)$ [5], where the $N$ are the level densities and $g$ the statistical weights for the upper ($u$) and lower ($l$) levels. We have used this parameter as an indicator of population inversion in this communication.

The present series of experiments was performed with different source geometries and with varying states of sooting of the hollow cathode. During emission spectroscopy, the He and Ne discharges were cycled as a function of $i_{dis}$ at constant pressures $P_{gas}$ ~1–2 mbar, while pressure variation at constant $i_{dis}$ was also studied for Ne. The timescale of cycling of $i_{dis}$ ensured that after each $i_{dis}$ step the stabilization time for the discharge was much greater than the time for data acquisition, which is ~ a few seconds with an Ocean Optics HR2000 spectrometer. It seems that population inversion is readily achieved for the singlet levels of He I while in the case of Ne there are some levels that show population inversion whereas some levels do not. Due to the limitations imposed by the spectrometer's lower wavelength limits only the singlet→singlet transitions of He I were chosen and these excited levels populate 2s $^1$S which has a mean lifetime of ~20 ms. This study may be interesting from the point of view of achieving population inversion in sources with interactive surfaces; regenerative soot is a prime candidate for such systems. In a related investigation [6] we have used the sooting discharges in regenerative mode to probe the possibility of its use as a VUV source of light. Regeneration of the soot implies the creation of sooting discharges where collisional excitation mechanisms involve not only the low energy electrons but the higher energy electrons as well as the metastable atoms.

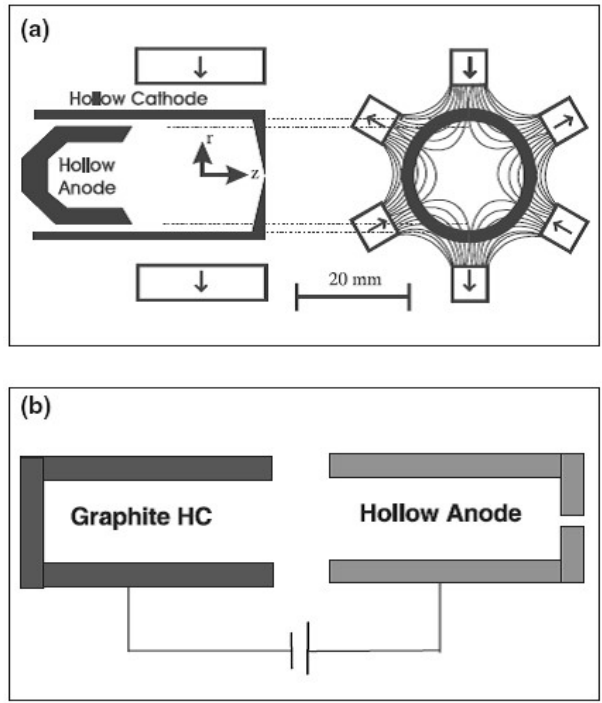

**Figure 1.** Two experimental set-ups of the source are shown. (*a*) shows the arrangement of the two inter-penetrating graphite cylinders as hollow cathode-HC and hollow anode-HA. A cross-sectional view of the hexapole, cusp magnetic field is also shown. (*b*) shows a simpler geometry of two hollow cylinders of graphite and iron facing each other and acting as the hollow cathode and anode in a geometry with large dimensions without the outer magnetic field.

## 2. Experimental

Figure 1 shows two designs of the graphite hollow cathode-HC sources used for the present study of the excited states of He I and Ne I in sooting discharges. Figure 1(*a*) shows the geometry that has already been used and described in [1]; it has two inter-penetrating, coaxial graphite cylinders. Any of these two HC cylinders can be used as the cathode. In addition, a hexapole magnetic field is also used in some of the experiments that we describe in this communication. The mode of operation with source geometry in figure 1(*a*) is to let the source be operated for a few hours at moderate discharge parameters, i.e. $V_{dis} \sim 0.5$–1 kV, $i_{dis} = 50 - 100$ mA, $P_g \sim 10^{-1}$ mbar. This ensures a well-sooted cathode surface from which electron as well as C cluster emission takes place during the discharge. The other geometry is shown in figure 1(*b*): this has two large cylinders of graphite and iron operated as HC and hollow anode, respectively. This arrangement does not use the magnetic field, and the inner cathode surface area is an order of magnitude larger than that in figure 1(*a*).

The emission spectroscopy of the discharges was done by two spectrometers; for the arrangements in figure 1(*a*) we used a synchronous motor driven Jobin Yvon monochromator with 1 Å resolution. Data acquisition takes 30 min for a full spectrum between 180–650 nm. After the initial calibration of the experimental set-up two kinds of data acquisition methods were employed: (i) a complete spectrum acquisition for a given set of experimental parameters, which required 30 min per spectrum; (ii) the monochromator is set on the desired wavelength for studying the cycling of $i_{dis}$ between the minimum and maximum values—this can be done instantaneously after the desired changes are done. An Ocean Optics spectrometer HR2000 was used in studies of the HC discharges in figure 1(*b*). These experiments needed a complete spectrum recording on a timescale of ∼1 s. HR2000 is directly coupled with the PC and we used mildly sooting discharges for observing the population inversion of the higher lying levels in He I and Ne I. The use of the analogue spectrometer (Jobin Yvon) implies the creation of a discharge that has to be operated for longer time durations, $10^3$ s, which inevitably becomes a well-sooted cathode surface, whereas the CCD based data acquiring spectrometer (Ocean Optics) can give the temporal evolution of the cathode surface being coated with soot. Therefore, using these two spectrometers we were able to study the population inversion among the support gases' excited levels as a function of the state of sooting of the hollow cathode surfaces.

## 3. Helium discharges in graphite HC

### 3.1. Well-sooted hollow cathodes

One of the parameters in the study of sooting discharges relates to the state of sooting, i.e. whether the discharge is predominantly carbonaceous or its support gas is the major contributor. This can be checked in graphite hollow cathodes where the initial discharge's emission spectrum normally has traces of $C_1$, $C_2$,... along with the dominant emission lines of the discharge support gas like He, Ne. The initiation of carbon's atomic and molecular species containing discharge alters the cathode's emissive surface by depositing a sooted layer which in turn becomes the cathode's effective surface. Mass spectrometry helps us to further establish whether the discharge has become predominantly sooted, i.e. when the mass spectrum shows a large range of C clusters emitted from such a discharge. This point has already been discussed in detail in [1]. Figure 2 presents two spectra of the level density of 3p $^1P^0$ calculated from the intensity of the line at $\lambda = 501.5$ nm.

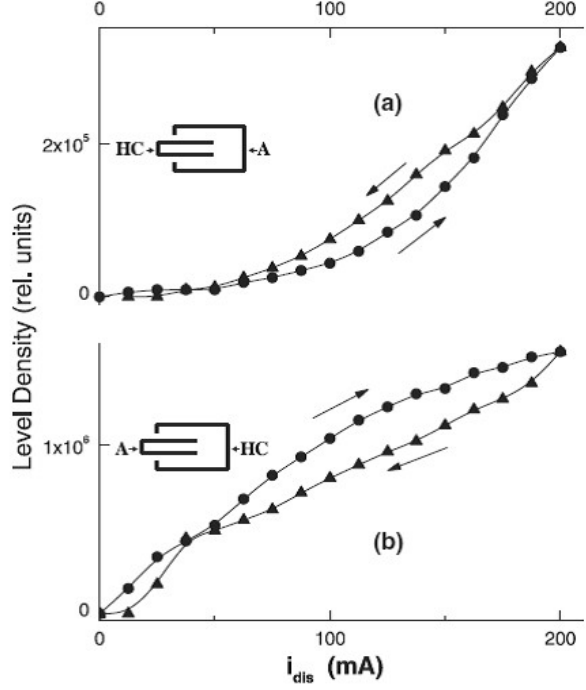

Figure 2. The emission spectra from He I's 3p $^1P^0$ level density are shown for the two geometrical arrangements of the source shown in figure 1(*a*). The 3p level density is calculated from the line at $\lambda = 501.5$ nm and plotted against the discharge current $i_{dis}$. This transition (with $\tau = 75$ ns) populates the long-lived metastable level 2s $^1S$ (with $\tau = 20$ ms). The insets in the two figures show the appropriate experimental geometry.

The two spectra are taken at constant $P_{He} \approx 1$ mbar as a function of $i_{dis}$ between 12.5 and 200 mA; the cycling sequence being $i_{dis} = 12.5 \rightarrow 200 \rightarrow 12.5$ mA. The level density of 3p is calculated using $I_{3p \rightarrow 2s} = N_{3p} \cdot h\nu_{3p \rightarrow 2s} \cdot A_{3p \rightarrow 2s}$, where $h\nu_{3p \rightarrow 2s} = (E_{3p} - E_{2s})$ is the energy difference between the two levels and $A_{3p \rightarrow 2s}$ the Einstein transition probability for spontaneous emission. This formula for the transition assumes the electron in the excited state to be a dipole oscillator and remain valid for the discharges in LTE equilibrium or non-LTE situations [7]. Figure 2(*a*) has the central cylinder as the HC while in 2(*b*) the outer one is the HC. The cycling of $i_{dis}$ shows hysteresis in figure 2(*a*) and not in figure 2(*b*). This situation is similar to the hysteresis seen and reported by Uzair *et al* (2003) [3] for the level densities of CI and CII.

Figure 3 is the levels diagram of those levels that participate in the transitions that are being discussed in this paper. Only the singlet levels are chosen and while six transitions are shown only three level densities will be presented and discussed. 3p, 3d and 4d level densities are calculated from the intensities of $\lambda = 501.5$ nm, $\lambda = 667.8$ nm and $\lambda = 492.1$ nm, respectively. The transitions from 3d and 4d populate 2p, which can either decay to 2s via a microwave transition $\lambda = 2058$ nm with $\tau = 500$ ns, or to the ground state with two VUV lines at $\lambda = 58.4$ nm, $\tau = 0.56$ ns and $\lambda = 53.7$ nm, $\tau = 1.8$ ns, respectively. The lifetimes of all the levels are also shown in the figure. All the three transitions that we have chosen for the investigation of the pattern of the cycling of singlet level densities of He I have one thing in common, namely all of them populate the metastable level 2s with $\tau = 20$ ms as opposed to the resonant transitions from 3p and 2p to the ground level with $\tau$ 1–2 ns.

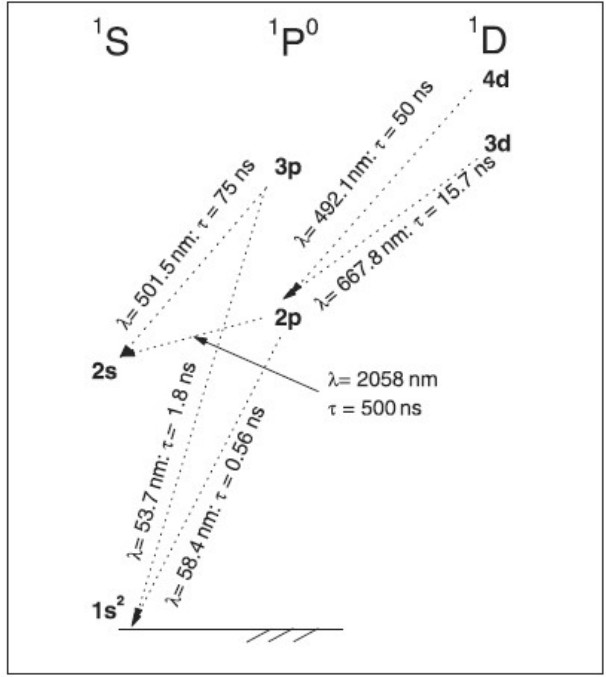

**Figure 3.** The energy level diagram of He I shows those representative transitions that are relevant for our discussion. Only the singlet–singlet transitions are shown that populate 2p and 2s levels. The respective wavelengths and the mean lifetimes of the transitions ($1/A_{u \to l}$) are shown. Out of the six transitions shown in the level diagram only three are observed and discussed in this communication. The two resonant ones at $\lambda = 58.4$ nm and $\lambda = 53.7$ nm are in the VUV and the third at $\lambda = 2058$ nm in the microwave region and thus not seen with our spectrometers.

### 3.2. Mildly-sooted hollow cathodes

The experimental set-up shown in figure 1(*b*) is used in this set of experiments. This arrangement has a cathode with an order of magnitude larger surface area as compared with the geometry for the well-sooted experiments shown in figure 1(*a*). The gas pressure, however, is the same as that for the case in figure 3, i.e. $P_{He} \approx 1$ mbar. In figures 4(*a*) and (*b*) the densities of the upper (4d) and lower (3d) levels are plotted as a function of $i_{dis}$. We call these the upper and lower levels since the two transitions at $\lambda = 492.1$ nm and $\lambda = 667.8$ nm have 2p as the common level for these transitions. The respective level densities are plotted as a function of the increasing and decreasing $i_{dis}$, so that if the level densities were to follow a Boltzmann distribution then one could extract an excitation temperature for the discharge. One can clearly see that the population inversion is taking place during the increasing and decreasing cycles of the discharge current. The non-LTE sooting discharge has provided the environment in which the singlet levels of He I show population inversion. Both the conditions of mild or well-sooted discharges introduce additional sources of excitation that may be responsible for the discharge that does not follow a Boltzmann distribution for the creation of the excited states by low energy electron collisional excitations alone. Population inversion is taking place and we may not use this set of transitions for the evaluation of the excitation temperature to estimate the electron temperature $T_e$ of the positive column of the He discharge.

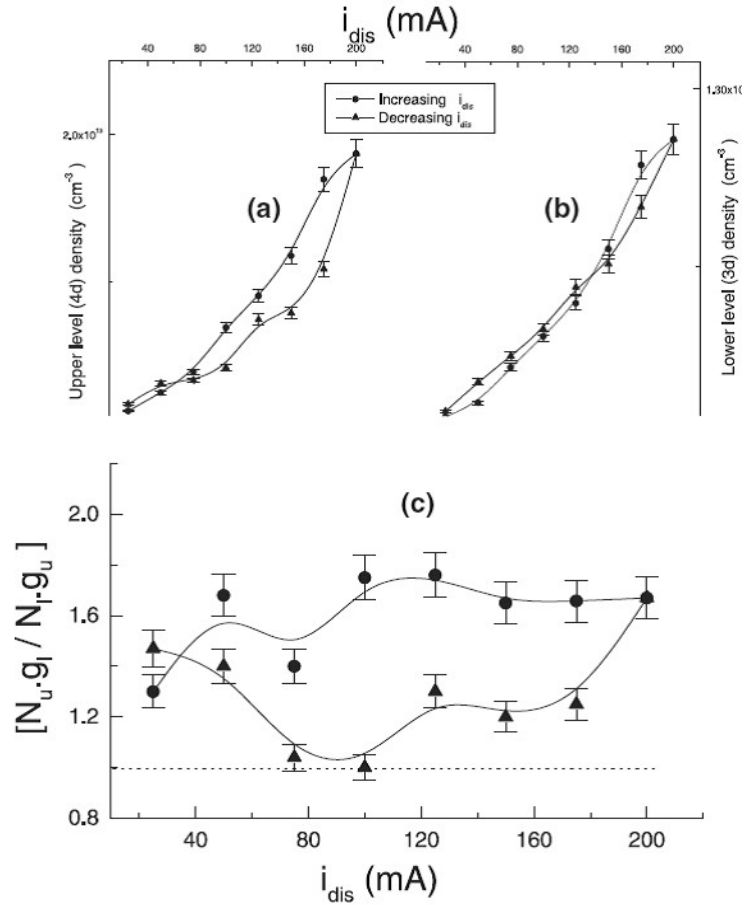

**Figure 4.** At constant pressure $P_{He} \approx 2$ mbar, the densities of the 4d and 3d levels of He I are shown in (a) and (b) for the geometry shown in figure 1(b). The level densities are plotted for the increasing and decreasing $i_{dis}$ from the transitions at $\lambda = 492.1$ nm and $\lambda = 667.8$ nm, respectively. These two transitions populate the same lower level 2p. In (c) the population inversion parameter $(N_u \cdot g_l / N_l \cdot g_u)$ is plotted as a function of $i_{dis}$. $(N_u \cdot g_l / N_l \cdot g_u) > 1$ for the increasing as well as the decreasing cycle of the discharge current between the minimum and maximum values.

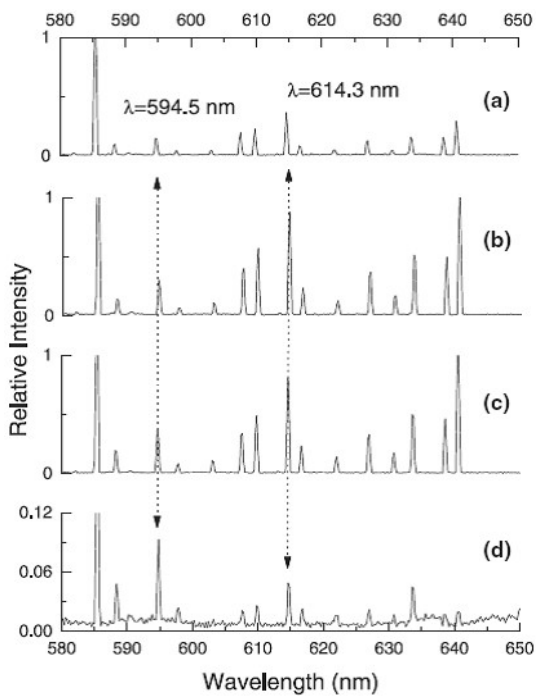

**Figure 5.** Using the geometrical arrangement of figure 1(a), Ne I emission lines are shown for increasing neon pressure from $P_{Ne} \approx 0.1$ mbar in (a) to 2 mbar in (b), 12 mbar in (c) and 20 mbar in (d). The relative intensities are plotted against the wavelength in the range 580–650 nm. In this range a large number of Ne I transitions are visible. The most noticeable feature of intensity versus $P_{Ne}$ is the population inversion clearly seen in the two transitions at $\lambda = 594.5$ nm and $\lambda = 614.3$ nm that fall to the same lower level $3s[1\frac{1}{2}]$. In (d) the intensity of the upper level is more than that from the lower level's transition.

# 4. Neon discharges in graphite HC

## 4.1. Well-sooted hollow cathodes

In figure 5 four spectra are shown from the neon discharge in well-sooted HC geometry of figure 1(*a*) with a hexapole, cusp magnetic field around the HC. The spectra are taken at constant $i_{dis}$ = 75 mA but the variable parameter is the neon gas pressure $P_{Ne}$. $P_{Ne}$ = 0.1, 2, 12 and 20 mbar for the spectra (*a*), (*b*), (*c*) and (*d* ), respectively. Without recourse to the detailed calculations, one can see that the relative intensities of the two marked emission lines at $\lambda$ = 594.5 nm and $\lambda$ = 614.3 nm that fall to the same lower level, are reversed as one gradually increases $P_{Ne}$ from 12–20 mbar. It must be emphasized that both these transitions populate the lowest metastable level of Ne I 3s [1 $\frac{1}{2}$] at 16.62 eV as shown in the level diagram of the four selected upper levels in figure 6.

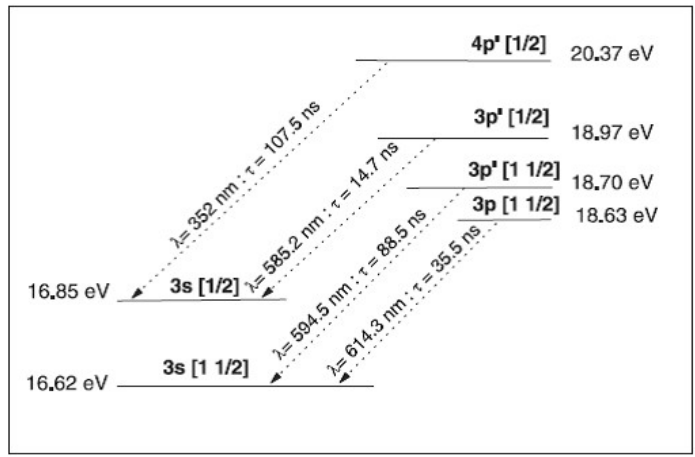

**Figure 6.** The energy level diagram of Ne I shows four of the representative transitions of the two sets that fall to the same lower levels. Two transitions each of the two sets are at $\lambda$ = 352.0 nm and $\lambda$ = 585.2 nm and $\lambda$ = 594.5 nm and $\lambda$ = 614.3 nm. The respective mean lifetimes of the transitions ($1/A_{u \rightarrow l}$) are shown.

There are a large number of Ne I transitions in figure 5 that show an interesting pattern of the variation of the level densities as a function of $P_{Ne}$ and one such set of transitions is at $\lambda$ = 352.0 nm (which is not shown in the figure but present in the total spectra) and $\lambda$ = 585.2 nm. Both these transitions fall to the same lower level 3s[ 1/2 ]. This set does not show population inversion in either the pressure increase or $i_{dis}$ cycling. The population inversion parameter ($N_u \cdot g_l / N_l \cdot g_u$) is calculated from ratios of two sets of level densities, i.e. 3p [1 $\frac{1}{2}$], 3p[1 $\frac{1}{2}$] and 4p [$\frac{1}{2}$], 3p[$\frac{1}{2}$]. This parameter is presented in figure 7(*a*) as a function of $P_{Ne}$, and as a function of $i_{dis}$ in figure 7(*b*). It can be seen that ($N_u \cdot g_l / N_l \cdot g_u$)<1 for the ratio of level densities of 4p [$\frac{1}{2}$] and 3p[$\frac{1}{2}$] plotted as 352/585.2 in figure 7. However, for the other set of levels, 3p [1 $\frac{1}{2}$] and 3p[1 $\frac{1}{2}$], shown as 594.5/614.3 in the graphs, there is a transition from non-population inversion up to 12 mbar pressure to a sudden inversion as the pressure is raised to 20 mbar. This effect is clearly visible in figure 5(*d* ) as well. ($N_u \cdot g_l / N_l \cdot g_u$) remains 1 for the entire range of $i_{dis}$ as shown in figure 7(**b**). All the results discussed above results are from a well-sooted discharge.

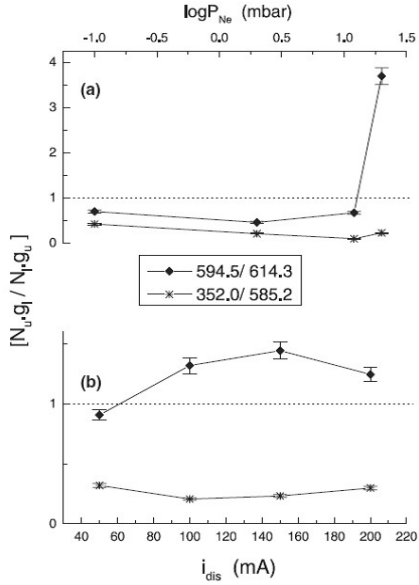

**Figure 7.** (*a*) The population inversion parameter ($N_u \cdot g_l / N_l \cdot g_u$) is plotted as a function of $P_{Ne}$ for the ratio of the two sets of transitions that, respectively, fall to the same lower levels. As seen in the figure 5 the $\lambda = 594.5$ nm and $\lambda = 614.3$ nm transitions show clearly the population inversion as the pressure is increased to 20 mbar. The other two transitions $\lambda = 352.0$ nm and $\lambda = 585.2$ nm do not show this trend. (*b*) shows the same set of transitions as a function of $i_{dis}$ and here too the ratio of $\lambda = 594.5/\lambda = 614.3$ shows inversion whereas $\lambda = 352.0/\lambda = 585.2$ does not.

## 4.2. Mildly-sooted Ne discharge

Ne discharges were also investigated in the mildly sooted discharges created in the experimental arrangement of figure 1(**b**). The results are shown for the set of levels that produce the two emission lines at $\lambda = 594.5$ nm and $\lambda = 614.3$ nm, as discussed earlier. Figure 8 is constructed to look into the cycling pattern of the upper (3p′) and lower (3p) level densities as a function of the discharge current. The two cycles of level densities are shown in figures 8(**a**) and (**b**). There is no hysteresis, as was seen in He I 3p level densities in figure 2(**a**), but the population inversion parameter $N_u \cdot g_l / N_l \cdot g_u$ when plotted in the inset shows a peculiar inversion trend that decreases as $i_{dis}$ is increased. The pressure of the discharge was maintained at 2 mbar.

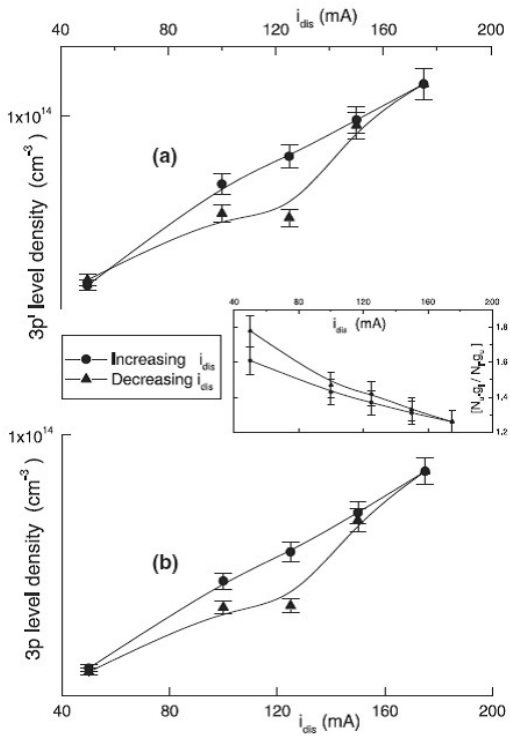

**Figure 8.** The densities of the 3p′ and 3p levels of Ne I are shown in (*a*) and (*b*) for the geometry shown in figure 1(*a*) at a constant pressure $P_{Ne} \approx 2$ mbar. The level densities are plotted for the increasing and decreasing $i_{dis}$ from the transitions at $\lambda = 594.5$ nm and $\lambda = 614.3$ nm, respectively. These two transitions populate the same lower level $3s[1\tfrac{1}{2}]$. In the inset the population inversion parameter ($N_u \cdot g_l / N_l \cdot g_u$) is plotted as a function of $i_{dis}$. ($N_u \cdot g_l / N_l \cdot g_u$) >1 for the increasing as well as the decreasing cycle of the discharge current between the minimum and maximum values, but increases while the discharge current is reducing.

## 5. Conclusions

We have studied the discharges of helium and neon in graphite hollow cathodes with interesting observations of hysteresis of the level populations of CI and CII when the discharge current was cycled, as in a recent study [3]. This prompted us to look at the level densities of all the discharge species especially those of the support gases, i.e. Ne and He. In multi-component discharges electron induced collisional excitations and ionizations are complemented by collisions between atoms and ions. In addition, the role of the surface also becomes important. The discharges that are in local thermodynamic equilibrium (LTE) provide level populations in the emission spectra of the neutral and ionized species that follow a Boltzmann distribution. This helps us to estimate the electron temperature $T_e$ of the discharge from the excitation temperature deduced from the set of transitions that fall to the same lower level. Whereas, transitions to the ground level are best suited for such estimates of $T_e$, other transitions to excited or metastable levels may provide such estimates. The situation in hollow cathode discharges, in general, is complicated by the existence of two regimes of electrons as pointed out in the introduction. By using the emission spectra one can only estimate the plasma column's excitation temperature; this was discussed in [1]. The existence of loose agglomerates of carbon clusters $C_x$ ($x \geq 2$) on the graphite hollow cathode's inner surface provides the additional interacting species as well as a surface layer that may participate in transforming the discharge's character. In this regard, we identify two categories of sooted layers on top of the original graphite surface; one that is mildly sooted and the other that has been covered by multiple layers of C clusters. The distinction between the two is based on the temporal history of the discharge, which includes the duration for which a particular discharge has been on, the type of the gas and the discharge power. A rule of thumb is that in the case of mildly sooting discharges $C_1$'s emission lines dominate the spectra while for well-sooted discharges the presence of $C_2$'s Swan (0,0) band at 516.5 nm is predominant.

We have employed two different types of spectrometers one with a slow data acquisition, but with higher resolution, while the other one records a chosen part of the spectrum at a higher speed (13 ms is the minimum time for a 2048 data point acquisition) with lower resolution. By optimizing the two instruments, we have been able to conduct experiments where the state of sooting of the discharge was an important parameter. We have shown in the results from the excited level densities of He I and Ne I that departures from LTE are prominent among the singlet levels of He I while population inversion takes place among some of the excited levels of Ne I.